# Optimizing hot electron harvesting at planar metal-semiconductor interfaces with titanium oxynitride thin films

Brock Doiron[A], Yi Li[A], Andrei Mihai[B], Stefano Dal Forno[A], Sarah Fearn[B], Lesley F. Cohen[A], Neil M. Alford[B], Johannes Lischner[B,C], Peter Petrov[B], Stefan A. Maier[D,A], Rupert F. Oulton[A]

[A] Department of Physics, Imperial College London, London, UK
[B] Department of Materials, Imperial College London, London, UK
[C] Thomas Young Centre for Theory and Simulation of Materials, Imperial College London, London, UK
[D] Nanoinstitut München, Chair in Hybrid Nanosystems, Faculty of Physics, Ludwig-Maximilians Universität München, München, Germany

## ABSTRACT

Understanding metal-semiconductor interfaces is critical to the advancement of photocatalysis and sub-bandgap solar energy harvesting where sub-bandgap photons can be excited and extracted into the semiconductor. In this work, we compare the electron extraction efficiency across $Au/TiO_2$ and titanium oxynitride/$TiO_{2-x}$ interfaces, where in the latter case the spontaneously forming oxide layer ($TiO_{2-x}$) creates a metal-semiconductor contact. Time-resolved pump-probe spectroscopy is used to study the electron recombination rates in both cases. Unlike the nanosecond recombination lifetimes in $Au/TiO_2$, we find a bottleneck in the electron relaxation in the TiON system, which we explain using a trap-mediated recombination model. Using this model, we investigate the tunability of the relaxation dynamics with oxygen content in the parent film. The optimized film ($TiO_{0.5}N_{0.5}$) exhibits the highest carrier extraction efficiency ($N_{FC} \approx 2.8 \times 10^{19} m^{-3}$), slowest trapping and an appreciable hot electron population reaching the surface oxide ($N_{HE} \approx 1.6 \times 10^{18} m^{-3}$). Our results demonstrate the productive role oxygen can play in enhancing electron harvesting and elongating electron lifetimes providing an optimized metal-semiconductor interface using only the native oxide of titanium oxynitride.

Plasmonic devices allow unprecedented control of light on the nanoscale[1] and highly-sensitive molecular detection[2] through the increased interaction between a conductor's free carriers and light via surface plasmon resonances. Although plasmonic modes decay on the order of tens of femtoseconds[3], much of the energy remains in excited carriers that relax ultimately through lattice interactions over picosecond timescales. Exploiting the energy that remains in these carriers has evolved into so-called 'hot-carrier' applications. For example, the use of a Schottky barrier to collect hot electrons (holes) into the conduction (valence) band of a semiconductor has underpinned the burgeoning field of sub-bandgap photodetectors and photovoltaic devices[4]. Another application involves the use of these energetic carriers in surface reduction and oxidation reactions for photocatalysis and solar water splitting[5]. Due to the low absorption of gold in the red and infrared, nanoparticles are needed to enhance the absorption but it comes at the cost of more expensive fabrication and the necessity of a range of particle sizes to best cover the solar spectrum. Transition metal nitrides provide a considerable advantage in such situations due to their strong broadband absorption[6] as well as the ability to tune their electronic and optical properties by varying deposition conditions.[7]

Titanium nitride (TiN), a ceramic with tunable stoichiometry, is known to have a high free carrier density such that it exhibits optical properties similar to gold in the visible and near-infrared regimes[8] but with significantly improved resilience to high temperatures.[7,9] Additionally, titanium nitride has been shown to achieve enhanced hot electron harvesting relative to gold,[10,11] and indeed it is reported that TiN has long-lived photoexcited carriers[12], but the physical origin of this phenomenon is poorly understood, as we show from theoretical considerations of the decay mechanisms in pure TiN. Better understanding of the material and its carrier decay dynamics hold the key to unravelling the underlying electronic processes taking place both within the material and during charge transfer to neighbouring materials. Although single crystalline TiN can be epitaxially grown on specific substrates[13], most sputtered TiN found in practical applications contains an unavoidable amount of oxygen due to nitrogen substitution at grain boundaries[14]. The physical properties of TiN are extremely sensitive to the substitution of oxygen within its lattice, enabling also the tuning of its optical response. Previously we have shown that titanium oxynitride films exhibit intermediate properties between titanium nitride and titanium dioxide, including the emergence of two tunable epsilon near zero (ENZ) points[15].

Here, we demonstrate how time-resolved pump-probe spectroscopy can be used to simultaneously investigate the electron dynamics in both metals and semiconductors as well as the specific dynamics associated with the interface between $Au/TiO_2$ (metal-semiconductor) and between $Au/SiO_2$ (metal-insulator). Using these as control samples, we show that titanium oxynitride (which is interfaced with its own semiconducting $TiO_{2-x}$ surface oxide layer[16]) exhibits a fundamentally different recombination mechanism than that at the Au/TiO$_2$ interface, showing carrier lifetimes beyond nanoseconds. By taking into account the results of density functional theory calculations and experimental material characterization (secondary ion mass spectroscopy and spectroscopic ellipsometry), we infer that the observed signal is due to an electron transfer process from the TiON into the TiO$_{2-x}$, where recombination takes place with the holes residing in the metal through trap-assisted processes. We introduce a kinetic model to discern the underlying physical processes, which illustrates how material composition influences the trapping and recombination of the extracted carriers. Variation of oxygen content in the underlying film allows tunability of free carrier densities at the surface by an order of magnitude. With our optimized film, we are able to demonstrate the presence of out of equilibrium hot carriers at the free surface, readily available to participate in surface chemical reactions.

Upon exposing TiN films to air it is recognised that a self-limiting nonstoichiometric semiconducting titanium dioxide (TiO$_{2-x}$ where x quantifies oxygen vacancies that may be present) self-limiting surface oxide forms, protecting the film against further oxidation or damage from external contaminants[17]. Conveniently, for sufficiently low oxygen vacancies this surface oxide is semiconducting but with a large bandgap between 3.4-3.6 eV for $0 < x < 0.3$.[18] Such metal-semiconductor interfaces are critical for many solar-based applications such as photovoltaics and solar water splitting due to the ability to separate energetic carriers that are excited with photon energies below the bandgap of the semiconductor. Separating these carriers before the electrons can thermalize with the lattice of the metal (typically on the order of several picoseconds[19]), allows for a much larger window for these carriers to be harvested or used in chemical reactions of nanoseconds or longer[20]. The several orders of magnitude difference of these lifetimes can allow for the resolution of individual contributions of the electrons remaining in the metal and those that have reached the semiconductor conduction band with a sub-picosecond laser.

To investigate the dynamics of the optically-excited electrons, we measure the time-resolved differential reflectivity using pump-probe spectroscopy with two <200 fs pulses. A 5 mW, 850 nm pump pulse excites carriers in the material and we then monitor the change in refractive index ($\Delta n$) by measuring the reflectivity of a time-delayed 150 $\mu$W, 1150 nm probe pulse. We begin by characterizing two control samples Au/SiO$_2$ and Au/TiO$_2$, the latter of which is known to exhibit a long-lived decay due to carrier separation at the Schottky barrier.[20] Figure 1a shows the time-resolved differential reflectivity of both samples on a semi-logarithmic plot. The Au/SiO$_2$ sample shows a rapid decay within 10 ps, which can be fitted to the two-temperature model[19], describing the interaction between high energy electrons and low energy phonons via exchange of thermal energy. This decay can be approximated as a biexponential function with decay lifetimes associated with electron-phonon and phonon-phonon scattering, herein represented as $\tau_{e-ph}$ and $\tau_{ph-ph}$, respectively. However, in addition to the Au response, the Au/TiO$_2$ sample exhibits a long-lived decay associated with free electrons in the conduction band of the TiO$_2$. The semiconductor contribution to the differential reflectivity is a direct measure of the electron harvesting efficiency as the signal is proportional to the free carrier concentration $N_{FC}$.[22] At normal incidence:

$$\left|\frac{\Delta R}{R}\right| = \frac{4}{n_0 - 1}|\Delta n| = \frac{4}{n_0 - 1}\frac{2\pi e^2}{n_0 m^* \epsilon_0 \omega^2} N_{FC}$$

where $e$ is the charge of an electron, $n_0$ is the unperturbed refractive index, $m^*$ is the effective mass and $\omega$ is the frequency of the probe beam.

By decomposing the individual contributions of the metal and semiconductor, we have a means of evaluating the electron extraction efficiency by purely optical means. Figure 1b shows the differential reflectivity of the Au/TiO$_2$ sample fitted to a sum of three exponential decay functions over the first 10 ps. Using the fitted data, the signal is decomposed into separate Au and TiO$_2$ signals showing that beyond 5 ps, the signal is dominated by the TiO$_2$ response. Using the maximum of the TiO$_2$ contribution, we can estimate the extracted carrier concentration to be $2.8 \times 10^{18}$ m$^{-3}$. In the inset of Figure 1b we show the fitted Au contribution along with the Au/SiO$_2$ measurement scaled by a factor of 0.4. The similarity of the two decays reinforces the validity of our decomposition. With our control samples well understood, we now look to compare the Au/TiO$_2$ behaviour with that of TiON/TiO$_2$. Since it is only the free electrons in TiO$_{2-x}$ injected from TiON that absorb the 1150 nm probe pulse[21], we know that we only measure the free electrons, which are capable of participating in chemical reactions. The two normal differential reflectivity signals are plotted alongside the parameters extracted from the fits in Figure 1c. There is a striking difference observed between the two samples, particularly beyond 200 ps where the recombination in the TiON sample appears to bottleneck. This difference at longer timescales suggests that the recombination process merits much more detailed material analysis.

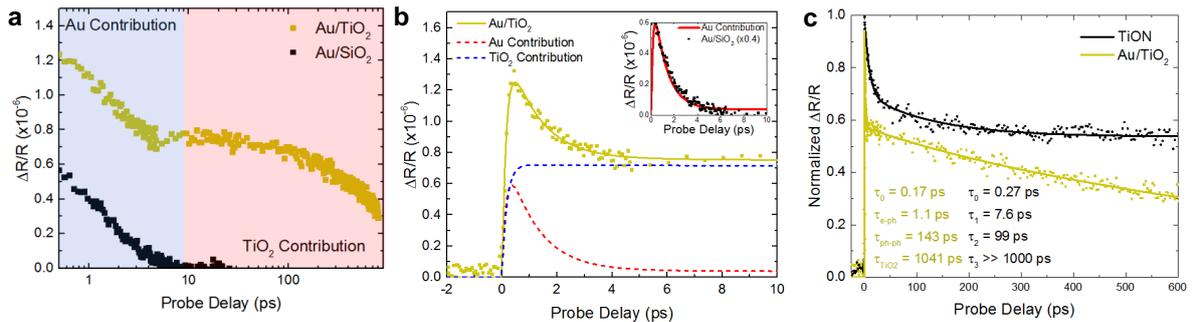

**Figure 1| Time-resolved pump probe spectroscopy to investigate dynamics at a metal-semiconductor interface.** (a) Semilogarithmic plot of the time-resolved differential reflectivity of Au/TiO$_2$ (metal-semiconductor) and Au/SiO$_2$ (metal-insulator). The Au response (blue shaded region) is clearly delineated from

the TiO$_2$ response (red shaded region) confirmed by the Au/SiO$_2$ control sample, which only has the Au response. (b) Using a combination of two-temperature model and metal-semiconductor recombination, the two contributions are separated showing long-lived electrons remaining in the TiO$_2$ conduction band. The inset shows the fitted Au contribution alongside the Au/SiO$_2$ sample (scaled by a factor of 0.4) showing very good agreement. (c) Differential reflectivity signals of TiON and Au along with the fitted lifetimes. Beyond 300 ps, the TiON signal is constant in contrast to the typical Schottky barrier recombination lifetime seen in the Au/TiO$_2$ suggesting a different recombination mechanism.

First we examine how the elemental composition varies along the depth of the films using time-of-flight secondary ion mass spectroscopy (ToF-SIMS). Using the composition of the material at the mid-point between the surface (t=0) and substrate (t=625) we determine the film shown in Figure 1 to be nearly stoichiometric TiO$_{0.2}$N$_{0.8}$ (Ti: 50%, N: 40%, O:10%, which we name TiON 10% to reflect the relative oxygen content of the film), shown in Figure 2a. The film displays a sharp increase of oxygen and decrease of nitrogen at the surface confirming the presence of interfacial TiO$_{2-x}$. Below that, there is a rapid onset of nitrogen content within the first 50 s of sputtering followed by a secondary slower increase of nitrogen until it stabilized around 250 s. The latter transitional layer is a result of oxygen diffusing into the film, and shows a decrease with depth due to the energy barrier for oxygen at the surface to penetrate into the bulk of the film[23,24]. It is thought that the randomness associated with the oxygen substitution has the potential to increase disorder at the interface and could result in a higher density of oxygen vacancies.[25] To investigate the absorptive properties of the film we use spectroscopic ellipsometry fitting to a Drude-Lorentz model with an oxide layer to determine the permittivity (Supplementary Section S1). Using this we calculate the absorption coefficient plotted in Figure 2b along with that of Au and the fitted Drude (free carrier) contribution to the TiON. As TiO$_{2-x}$ only absorbs photons above its bandgap energy (wavelengths below 400 nm) shown by the dark red curve in Figure 2b, we conclude that the absorption at 850 nm is via the free carrier absorption in the underlying TiON 10% film.

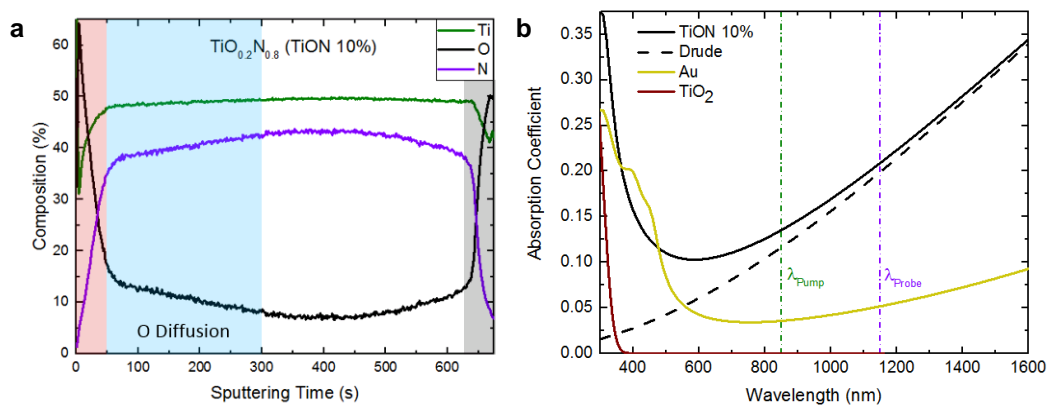

**Figure 2| Titanium oxynitride material characterization.** (a) Time-of-flight secondary ion mass spectroscopy (ToF-SIMS) characterization of the elemental composition profile of the film determined to be TiO$_{0.2}$N$_{0.8}$, referred to as TiON 10%. The TiO$_{2-x}$ surface oxide (red shaded region) and SiO$_2$ substrate (grey shaded region) form the two boundaries of the TiON 10% film. Oxygen diffusion into the film is clearly visible below the surface oxide (blue shaded region). (b) Absorption spectra of TiON and Au measured by spectroscopic ellipsometry and fitted to a Drude-Lorentz model along with the pump (green) and probe (purple) wavelengths used in this study. The

free electron (Drude) contribution of TiON is shown with the dashed line and $TiO_2$ is shown as a reference to show that there is no surface oxide absorption in the TiON film.

As we are looking at a novel material system, little is known about the interactions between electrons and phonons within the material. The previous demonstration of a long-lived differential reflectivity signal in TiN films was interpreted in terms of weak electron-phonon coupling.[12] The temperature-dependent electron and phonon coupling of TiN can be modelled using ab-initio DFT+U calculations, which we present in the Supplementary Section S5. Remarkably, the results reveal that the pure TiN electron-phonon coupling constant is two orders of magnitude greater than that of Au. The corresponding electron-phonon lifetimes are shown in Figure 3a, which are on the order of 100 fs or faster. The inset of Figure 3a compares the calculated electron-phonon lifetimes of TiN and Au and although this agrees very well with the calculated $\tau_{e-p}$ of our control sample, the lifetimes of pure TiN cannot explain the long experimentally observed lifetimes. Furthermore, as timescales this fast are below the temporal resolution power of our setup, the implication is that the entire observed signal must originate from the refractive index change associated with the occupation of the conduction band states in the $TiO_{2-x}$ surface oxide layer.

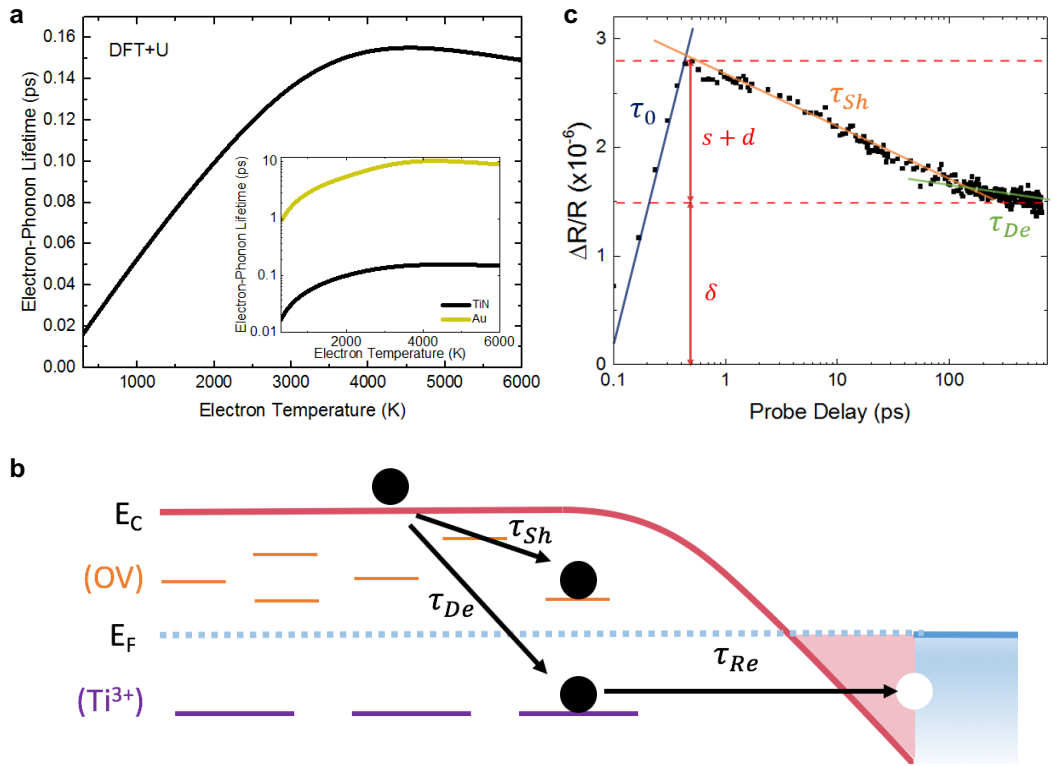

**Figure 3| Theoretical investigation of the TiON-$TiO_{2-x}$ interface and the trap-assisted recombination model.** (a) Ab initio DFT+U calculations of the electron-phonon scattering time of TiN. The strong electron-phonon coupling of TiN results in thermalization with the lattice within the resolution of our system. Inset shows a comparison of calculated Au and TiON lifetimes with Au showing two order of magnitude longer lifetimes. (b) Proposed trap-assisted recombination model at the TiON-$TiO_{2-x}$ interface with the associated lifetimes observable in our measurements associated with shallow trap occupation ($\tau_{Sh}$), deep trap occupation ($\tau_{De}$) and recombination ($\tau_{Re}$). (c) Experimental data showing the shallow and deep trap occupation lifetimes along with the exponential

rise time of the signal ($\tau_0$) and the amount of energy lost to occupying traps ($s + d$) and the amount remaining in the free carriers ($\delta$).

Since both electron injection into $TiO_{2-x}$ and electron relaxation in TiON occurs within the pulse width of the pump beam, what is required here is a realistic description of the subsequent relaxation and recombination channels of free carriers in $TiO_{2-x}$. We propose that the recombination occurs via trap states in the $TiO_{2-x}$ and at the interface with TiON, and that the recombination is slowed because of the saturation of said states and slow detrapping times. In $TiO_2$ there are two sources of trap states: oxygen vacancies (OV) forming traps close to the conduction band edge (shallow traps) and $Ti^{3+}$ forming deep traps within the bandgap. Similar to what is seen in the dye-sensitized titanium dioxide[26,27], we observe the occupation of the trap states occurring faster than recombination back into the metal until a quasi-equilibrium is reached when the trap states are fully occupied. Following this, the subsequent decay is known to be on the order of nanoseconds to milliseconds as it follows the rate of detrapping to recombine with the hole remaining in the metal. Figure 3b shows a schematic of the proposed recombination model and the associated lifetimes as is typically associated with shallow trapping ($\tau_{Sh}$) and deep trapping ($\tau_{De}$).[26,28] Figure 3c shows a semilogarithmic plot of the TiON 10% data with the proposed lifetimes clearly delineated. In addition we show $\tau_0$, which is a simple fit to the exponential rise of the signal associated with both the overlap of pump and probe pulses convolved with the rise associated with the occupation of the conduction band states.

The density of free electrons in the conduction band is described by $N_{FC}(t)$ and decays as the free electrons occupy the shallow ($N_{Sh}(t)$) and deep ($N_{De}(t)$) traps. Thus the temporal evolution of the free carriers is described by: $N_{FC}(t) = N_{FC}(0) - N_{Sh}(t) - N_{De}(t)$. The rate of occupation of these trap states is proportional to the availability of trap states, decreasing with increasing occupation. To quantify this, we describe maximum occupation values, S and D, for shallow and deep traps, respectively. We then denote the rates of proportionality for the shallow and deep states as $k_{Sh}$ and $k_{De}$ leading to the following two differential equations:

$$\frac{d}{dt}N_{Sh}(t) = k_{Sh}(S - N_{Sh}(t)) \qquad \frac{d}{dt}N_{De}(t) = k_{De}(D - N_{De}(t))$$

This has a simple solution, resulting in the biexponential decay:

$$\frac{n_{Free}(t)}{n_{Free}(0)} = (1 - s - d) + s\, e^{-t/\tau_{Sh}} + D e^{-t/\tau_{De}}$$

where $s = S/n_{Free}(0)$ and $d = D/n_{Free}(0)$. We denote the constant value $1 - s - d$ as $\delta$, which characterizes the residual electron occupation after the trap states are occupied. With the relationship between the free carrier concentration and differential reflectivity already established above, we can now relate the decays observed in TiON to the underlying physical processes.

In TiON 10% we observe the shallow trapping to be relatively quick compared to pure TiO$_2$, $\tau_{Sh} = 7.6 \pm 0.8\ ps$, compared to 29.8 ps in pure TiO$_2$[28] suggesting a significant amount of oxygen vacancies in this film. Furthermore, the fitted deep trap lifetime $\tau_{De} = 99 \pm 12\ ps$ is lower than pure TiO$_2$ (471 ps[28]) but is approximately the same as metal-doped TiO$_2$.[28] Thus, this is consistent with the increase in deep traps (Ti$^{3+}$) that have not reacted with oxygen. In order to explore the model further, we study three additional TiON films with a systematically increasing oxygen content: TiON 15%, TiON 25%, and TiON 40%. We anticipate that the increased oxygen content in the underlying TiON films will react with the Ti$^{3+}$ ions and decrease both shallow and deep traps resulting in an increase to the observed lifetimes. Figures 4a-c show the oxygen content profile of each of the three additional films measured by ToF-SIMS and the associated differential reflectivity signals compared to the TiON 10% and Au/TiO$_2$ samples. By increasing the oxygen content slightly to 15% (Figure 4a), we observe a slightly larger differential reflectivity, but similar trapping lifetimes ($\tau_{Sh} = 5.7 \pm 0.8\ ps, \tau_{De} = 83 \pm 12\ ps$). This is to be expected as we still observe oxygen diffusion into the film, suggesting a porous surface oxide (shaded region).

When increasing the oxygen further to 25% we observe a distinct change in behaviour of both the oxygen profile and differential reflectivity as seen in Figures 4b. We observe uniform oxidation throughout the TiON film suggesting the TiO$_{2-x}$ formed is uniform and effectively blocks the further diffusion of oxygen into the film. This is also reflected in a substantial increase in the differential reflectivity signal, associated with more efficient electron extraction and the extended lifetimes of the TiON 25% ($\tau_{Sh} = 16.7 \pm 3.4\ ps, \tau_{De} = 324 \pm 70\ ps$), which now approach the values for pure TiO$_2$. This is a direct demonstration of using the oxygen content in TiON to tune the electron relaxation dynamics in the adjacent semiconductor to make it more favourable for hot electron applications. In this case we also observe a distinct ultrafast decay not present in either the TiON 10% and TiON 15%, which will be analyzed in detail in the subsequent section. The heavily oxidized TiON 40% exhibits similarly long lifetimes ($\tau_{Sh} = 12.3 \pm 4.6\ ps, \tau_{De} = 237 \pm 130$ ps) as well as the ultrafast peak. However, the magnitude of the differential reflectivity signal is weaker than that of TiON 25% due to the less metallic behaviour. In Figure 4d we show the estimated harvested carrier concentration using the fitted differential reflectivity data measured for each of the films. We estimate the extracted carrier concentrations to be: $N_{FC}^{10\%} = 1.0 \times 10^{19}\ m^{-3}, N_{FC}^{15\%} = 1.4 \times 10^{19}\ m^{-3}, N_{FC}^{25\%} = 2.8 \times 10^{19}\ m^{-3}, N_{FC}^{40\%} = 1.3 \times 10^{19}\ m^{-3}$, for the four TiON films. The superiority of TiON 25% is clear in that it is the most efficient interface to harvest electrons and the harvested electrons remain in the semiconductor conduction band for longer compared to the other films.

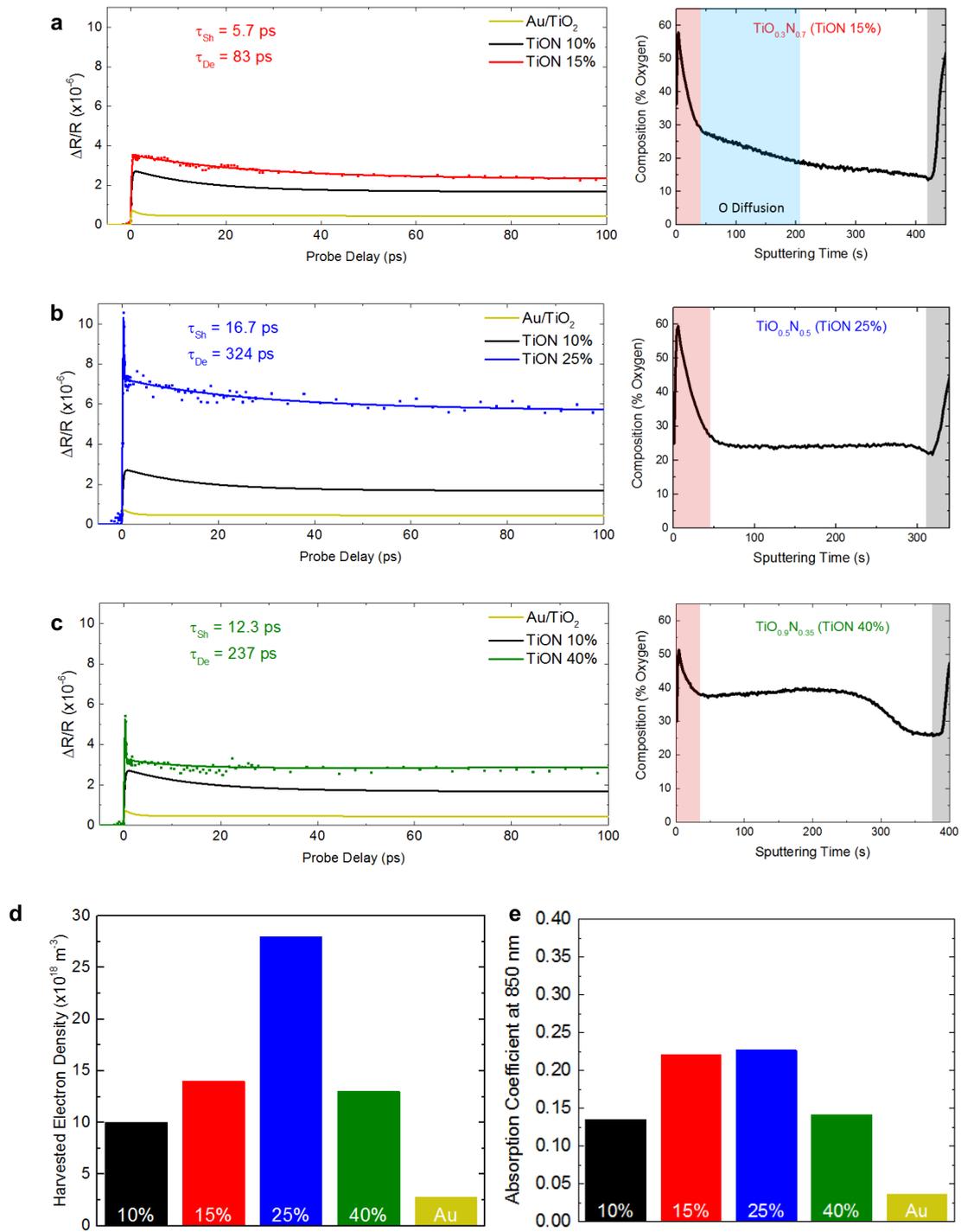

**Figure 4| Optimization of hot electron injection by tailoring the oxygen content in TiON thin films.** (a-c) Differential reflectivity following a 5 mW pump pulse (left) and oxygen composition (right) measured with ToF-SIMS for films with increasing oxygen content TiON 15% (a), TiON 25% (b), and TiON 40% (c). Each film exhibits a higher differential reflectivity than both the Au/TiO$_2$ and TiON 10% films over the entire temporal range. With sufficiently high oxygen included in the TiON films (25% and 40%) no post-deposition oxygen diffusion is detected as is seen in the TiON 10% and TiON 15% films. This more uniform interface results in more energetic electrons reaching the TiO$_{2-x}$ interface and the emergence of an ultrafast peak. (d) Maximum free carrier concentration in the TiO$_{2-x}$ determined via the fitting of differential reflectivity measurements. (e) Absorption coefficient at the pump wavelength (850 nm) for each film. It is clear that the strong enhancement in electron harvesting observed in TiON 25% cannot be explained by a stronger

Figure 4e shows the absorption coefficient for each of the materials at 850nm. The advantages of the TiON 25% film cannot be explained simply by an increase in absorption, which is comparable to TiON 15%. Furthermore, the significant increase in absorption of TiON 15% over TiON 10% only results in a modest increase in differential reflectivity. Thus, the injection behaviour between these films must differ in a way that favours transfer into the semiconductor. We posit that the low-defect interface of TiON 25% and 40% facilitates the direct injection of hot electrons into the conduction band of the $TiO_{2-x}$, which subsequently relax to the conduction band minimum via electron-phonon scattering. To investigate this injection behaviour more closely we have fit the ultrafast peak, as a Gaussian-shaped contribution, shown as the shaded regions of TiON 25% in Figure 5a and 5b. Using the height of the fitted Gaussian, we estimate the hot carrier concentration in the oxide layer to be $1.6 \times 10^{19} \, m^{-3}$. We estimate the lifetime, $\tau_{CB}$, of this population as the full-width-half-maximum (FWHM) of the fitted Gaussian shown schematically in figure 5b. Furthermore, the pump power dependence of $\tau_{CB}$ shown in Figure 5c indicates dynamics that depend on the electron temperature. It is known that hot electron effects show distinctive temperature-dependent relaxation dynamics[29] due to the relationship with volumetric heat capacity and electron-phonon coupling (Section S5).[29] This power-dependent behaviour is not observed in the subsequent two lifetimes ($\tau_{Sh}$ and $\tau_{De}$) as the electrons have lost their excess energy and remain at the conduction band minimum (Supplementary Section S4). It should be emphasized that this is not possible in the more metallic TiON 10% and TiON 15%, as the high disorder of the interface scatters the energetic electrons limiting transfer to the conduction band.[30]

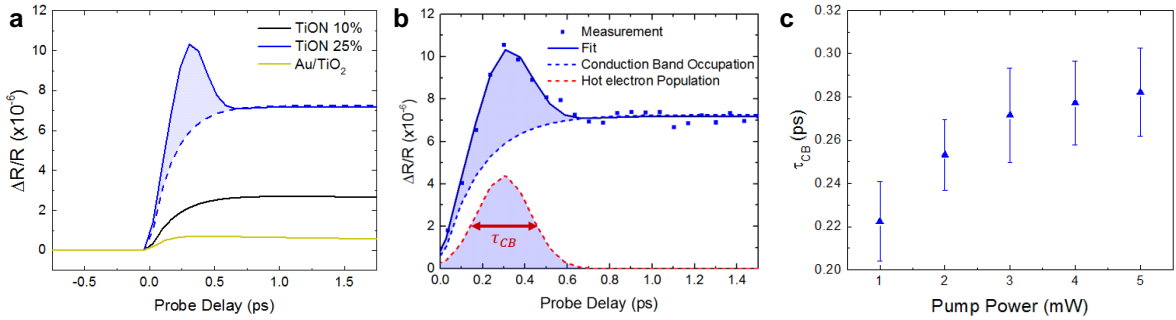

**Figure 5| Hot electron dynamics in the surface oxide of TiON 25%.** (a) Differential reflectivity measurements of Au/TiO$_2$, TiON10% and TiON 25% over the first 1.75 picoseconds. TiON 25% exhibits an additional ultrafast peak (shaded region) attributed to hot electrons in the TiO$_{2-x}$ relaxing to the conduction band minimum. (b) Decomposition of the TiON 25% signal into conduction band electron occupation (blue dashed line) and hot electron population (red-dashed line). The lifetime of the hot electron distribution ($\tau_{CB}$) describes the electron-phonon scattering time in the TiO$_{2-x}$. It is described by the fitted full-width half-maximum (FWHM) of the signal showed by the red arrow. (c) Power-dependence of $\tau_{CB}$, showing a slower relaxation with increasing power (and thus increased electron temperature) characteristic of a hot electron population.

In this work, we have closely examined the critical role that oxidation plays in hot carrier separation of titanium oxynitride thin films. We introduce an interfacial charge transfer model that provides a framework for describing the interplay between the strongly absorptive metallic layer and the omnipresent surface oxide layer. Hot electron injection into the conduction band of the surface titanium dioxide layer is achieved by engineering the interfacial oxygen defect states. Reducing the density of the traps in the surface oxide and at the interface also slows the energy loss over the first few hundred picoseconds. The subsequent recombination with the metal occurs on the timescale of nanoseconds or longer, confirming that we have trap-mediated carrier separation. The partially-oxidized titanium oxynitride film (TiON 25%) shows the most promise for use in hot carrier applications as it exhibits much more efficient carrier separation (as well as an estimated hot carrier concentration of $1.6 \times 10^{18}\ m^{-3}$ at the interface) and retains the greatest portion of its initial energy over the time period measured. Our work affirms the indispensable role that titanium oxynitride can play in the future development of plasmonic and hot carrier applications. Such a hybrid material system has improved versatility, using only its natural oxidation tendencies to achieve efficient extraction of electrons over a wider range of energies than is possible with comparable Schottky barrier systems.

**ACKNOWLEDGEMENTS**

We acknowledge support from the Engineering and Physical Sciences Research Council (EPSRC) Reactive Plasmonics Programme (EP/M013812/1), Lee-Lucas Chair in Physics, and the Henry Royce Institute made through EPSRC grant EP/R00661X/1. S.D.F. and J.L. acknowledge support from EPRSC under Grant No. EP/N005244/1 and also from the Thomas Young Centre under Grant No. TYC-101. Via J.L.'s membership of the UK's HEC Materials Chemistry Consortium, which is funded by EPSRC (EP/L000202), this work used the ARCHER UK National Supercomputing Service.

**EXPERIMENTAL SECTION**

Deposition

Titanium nitride films are deposited using a reactive RF magnetron sputtering from a titanium target in a $N_2$/Ar (30% $N_2$) plasma on fused quartz substrates. 50nm of each TiON 10% and TiON 25% was deposited at high temperature (600°C). For the TiON 10% oxygen contamination during growth was minimized by (<$5*10^{-9}$mbar O2 partial pressure) running a 1hr Ti pre-sputter (Ti is known to be a good O2 getter). A shorter pre-sputter for the 100nm TION 15% was done in order to get an intermediary O2 residual level in the sputtering chamber and the film was grown at high temperature (600°C). No pre-sputtering was done for the deposition of the TiON 25% film. More information on the growth methods can be found in reference 19.

Characterization

**Time-of-Flight Secondary Ion Mass Spectroscopy**

Using an ION-TOF TOF.SIMS 5 instrument a focused ion beam of $Bi_1^+$ ions are used to ablate the sample at a fixed power. The charged ions ejected from the surface are then collected and analysed based on their mass-to-charge ratio to determine the constituent molecules.

**Spectroscopic Ellipsometry**

Using a variable-angle JA Woollam VASE ellipsometer, the optical properties of the films used in experiments were determined using a Drude-Lorentz model. Fitting was performed using a Levenberg-Marquardt algorithm to minimize the mean squared error (MSE). The fitted parameters of each film are presented in Supplementary Section S1.

Differential Reflection Measurements

Using a Chameleon Ultra II Ti:Sapphire laser, an 850 nm pump pulse with temporal width below 200 fs is generated and a proportion used to produce a lower energy 1150 nm probe pulse using an optical parametric oscillator (OPO). The power of the pump pulse varied between 1 and 5 mW, but the probe is fixed at 125 $\mu$W. A motorized stage on the probe line allows for the delay between the two pulses to be controlled in steps as small as 30 fs. Using an mechanical chopper to modulate the pump beam and a photodetector and lock-in amplifier on the probe beam, time-resolved differential reflection measurements are extracted directly. A bi-exponential decay is used to Fitting is done using the Levenberg-Marquardt Algorithm and the quality of the fit is determined using the adjusted $R^2$ value, explained in Supplementary Section S2.

# SUPPLEMENTARY INFORMATION

S1 Titanium Nitride Film Ellipsometry

Spectroscopic ellipsometry data is fitted to a Drude-Lorentz oscillator with a $TiO_2$ surface layer. The model parameters and oxide thickness are fitted using the Levenberg-Marquardt Algorithm as described in Supplementary Section S3. The epsilon-near-zero (ENZ) wavelength indicates the crossover wavelength from dielectric to metallic behavior and is the point where the real part of the permittivity crosses from positive to negative. This gives an intuition into the operation range of surface plasmon polariton (SPP) and localized surface plasmon resonance (LSPR) applications. For the most oxidized film (TiON 40%), it remains positive over the entire wavelength range in question and so no ENZ point is determined. Next, the energies of the two interband transitions are shown, agreeing well with previous estimates for titanium nitride[1], except for the TiON 40% film. This suggests that the inclusion of oxygen in TiON 10%, 15% and 25% does not have a notable effect on the band structure. The high-frequency permittivity ($\epsilon_\infty$) is also given, accounting for high-frequency modes not included explicitly as Lorentz oscillators. The unscreened plasma frequency (without the Coulomb screening effects of the valence electrons) is given in terms of electron-volts (eV). This is the best indicator of the metallic behavior of the material as is proportional to the square root of the conduction electron density (N) through the following expression:

$$E_P^{US} = \hbar \sqrt{\frac{Ne^2}{m^*\epsilon_0}}$$

where $e$ is the charge of the electron, $m^*$ the electron effective mass and $\epsilon_0$ the vacuum permittivity. Finally, the Drude loss term, which characterizes the time between electron collisions, is given in femtoseconds accounting for electron scattering from impurities, phonons and other electrons.

|  | $TiO_2$ Thickness (nm) | ENZ (nm) | Interband 1 (eV) | Interband 2 (eV) | $\epsilon_\infty$ | $E_P^{US}$ (eV) | Drude Loss Term (fs) | MSE |
|---|---|---|---|---|---|---|---|---|
| TiON 10% | 6.4 | 495 | 5.61 | 3.96 | 1.82 | 5.7 | 9.4 | 3.3 |
| TiON 15% | 6.8 | 565 | 5.60 | 4.07 | 2.14 | 5.3 | 4.0 | 6.8 |
| TiON 25% | 10.2 | 625 | 5.62 | 3.75 | 2.36 | 5.0 | 2.7 | 2.9 |
| TiON 40% | 4.9 | N/A | 4.34 | 1.17 | 3.02 | N/A | 2.6 | 9.8 |

*Table S1 | Drude-Lorentz fitting parameters. The fitted parameters of the spectroscopic ellipsometry measurements to a Drude Lorentz model with two Lorentz oscillators fitted using a Levenberg-Marquardt algorithm to minimize the mean squared error (MSE).*

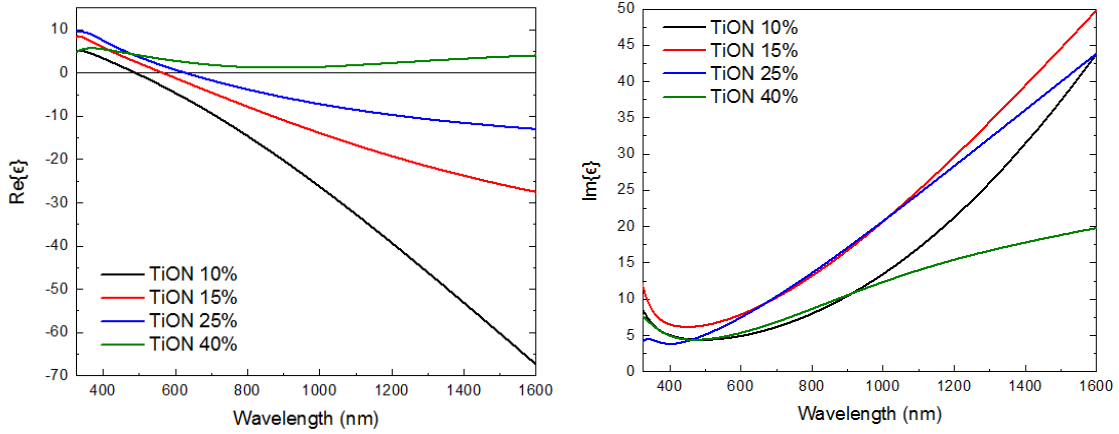

*Figure S1 | Optical properties of TiON thin films. The real (left) and imaginary (right) parts of the dielectric permittivity described by the Drude-Lorentz fitting parameters above.*

[1] Patsalas, P., Kalfagiannis, N. & Kassavetis, S. Optical properties and plasmonic performance of titanium nitride. *Materials (Basel).* **8,** 3128–3154 (2015).

S2 Evaluation of the Fit Accuracy

The Levenberg-Marquardt Algorithm is used to fit to fit both the Drude-Lorentz and electron kinetic models to the experimental measurements. In the electron kinetic model, five fitting parameters are used: the electron injection rise time ($\tau_0$), the shallow trap occupation time ($\tau_{Sh}$), the deep trap occupation time ($\tau_{De}$) and a the weights of the two trap states ($s$ and $d$). Thus the intention is to fit the function $f(t; \tau_0, \tau_{Sh}, \tau_{De}, s, d) = f(t, \boldsymbol{\beta})$ to the experimental data $y(t)$ by varying the parameters $\boldsymbol{\beta} = (\tau_0, \tau_{Sh}, \tau_{De}, s, d)$ minimizing the following first-order expression:

$$S(\boldsymbol{\beta} + \boldsymbol{\delta}) = \sum_i^m \left( y(t_i) - f(t_i; \boldsymbol{\beta}) - \nabla_\beta[f(t_i; \boldsymbol{\beta})]\boldsymbol{\delta} \right)^2$$

To evaluate the quality of the fit, the adjusted $R^2$ value is used to account for the number of fitting parameters ($p$) used. This is done to prevent overfitting, as the adjusted $R^2$ value only increases when including an additional parameter if it leads to a better fit than what is expected by random chance. [4]

$$R_{Adj}^2 = \frac{(1-R^2)(m-1)}{m-p-1}$$

Where R² is calculated using the standard formula:

$$R^2 \equiv 1 - \frac{\sum_i (y_i - f_i)^2}{\sum_i (y_i - \bar{y})^2}$$

The uncertainty on the fitted parameter is assumed to be one standard deviation on either side of the fitted value, giving a relatively high certainty (68.2%) that the true value is within the range stated.

*[4] Theil, Henri. Economic Forecasts and Policy. Holland, Amsterdam: North (1961).*

S3 Ab Initio Calculations of Hot Carrier Thermalization in Titanium Nitride

As the thermal electrons heat via electron-electron interactions, there are more collisions between the thermal electrons and the lattice vibrations. This is associated with a concurrent heating of the lattice and cooling of the electron system via the electron-phonon scattering events. The coupling between electrons and phonons can be described by an electron temperature-dependent coupling parameter G. Lin et al. [6] derived the following expression for an arbitrary density of electronic states, $\rho(\epsilon)$ and electron distribution function $f(E, T_e)$ as above:

$$G(T_e) = -\frac{\pi k_B \lambda \langle \omega^2 \rangle}{\hbar \rho(E_F)} \int \rho^2(\epsilon) \frac{\partial f(E, T_E)}{\partial E} \partial E$$

with $\lambda \langle \omega^2 \rangle$ being the electron-phonon mass enhancement parameter at the second moment of the phonon spectrum [6]. Neglecting the interaction between the nonthermal electrons and the phonons as the heating of the lattice is negligible in the short lifetime of the energetic electrons, we perform density-functional theory (DFT) calculations to determine the dependence of the electron-phonon coupling strength on the electron temperature in stoichiometric titanium nitride. Figure S5-1a shows the calculated band structure density of states and the associated temperature-dependent electron-phonon coupling parameter, G, is shown in Figure S5-2a. To characterize the corresponding changes in temperatures, it is imperative to understand the electron and lattice heat capacities defined by:

$$C_e(T_e) = \int_{-\infty}^{\infty} dE\, \rho(E) E \frac{\partial f(E, T_e)}{\partial T_e} \qquad C_l(T_l) = \int_{0}^{\infty} dE\, P(E) E \frac{\partial n(E, T_l)}{\partial T_l}$$

where $P(E)$ is the phonon density of states and $n(E, T_l)$ the Bose occupation factor for the phonon system. Figures S5-1b and S5-1c show the temperature dependences of the electronic and lattice heat capacities respectively. We then estimate the electron-phonon lifetime using the expression:

$$\tau_{e-p} \approx \frac{1}{G}\left(\frac{C_e C_l}{C_e + C_l}\right)$$

and plot the electron-temperature dependence in Figure S5-2b shown to be under 200fs even at higher temperatures.

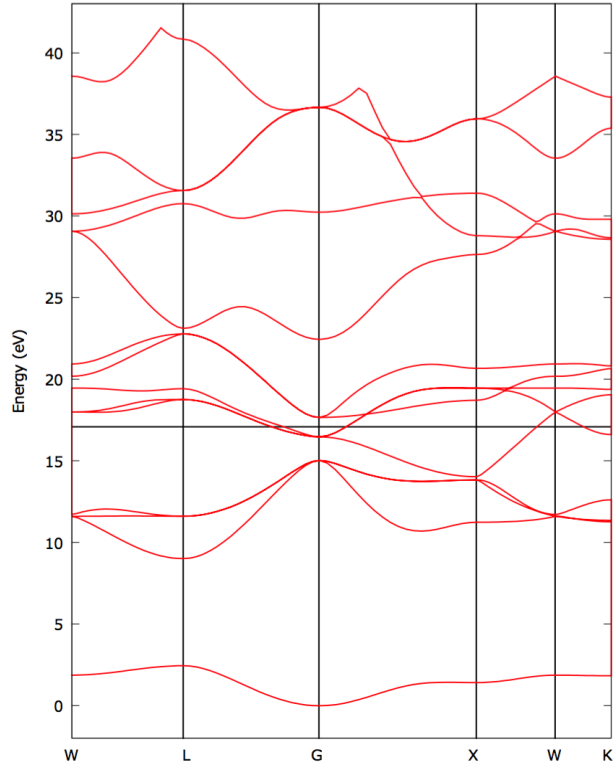

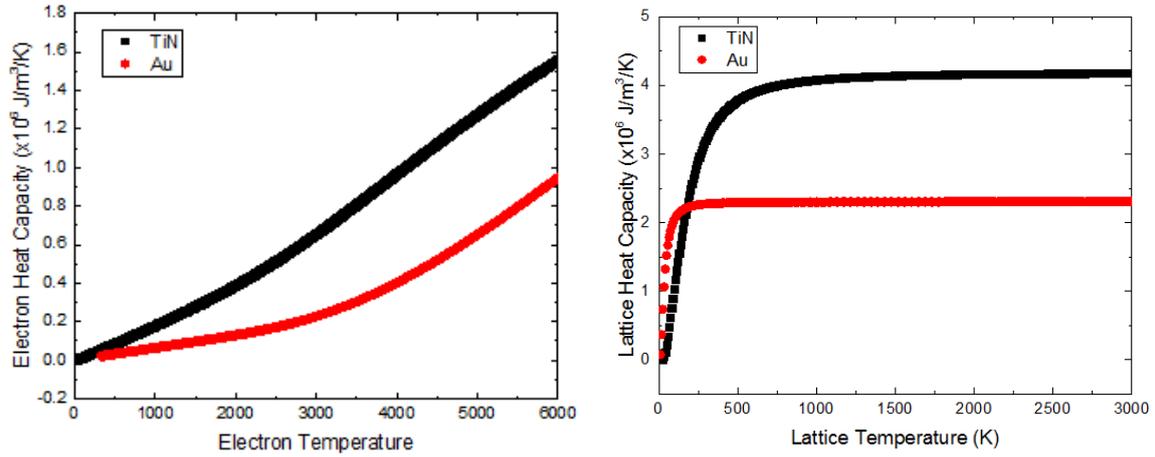

*Figure S3-1 | Density functional theory (DFT) calculations of electronic and thermal properties of TiN and Au*. *a*, electronic band structure and density of states of titanium nitride whose conductive properties are associated with the intersection of d-band states with the Fermi level. *b,* electronic heat capacity versus electron temperature showing a two linear regions with slightly different slopes. *c,* lattice heat capacity versus lattice temperature showing an increase even above 500K.

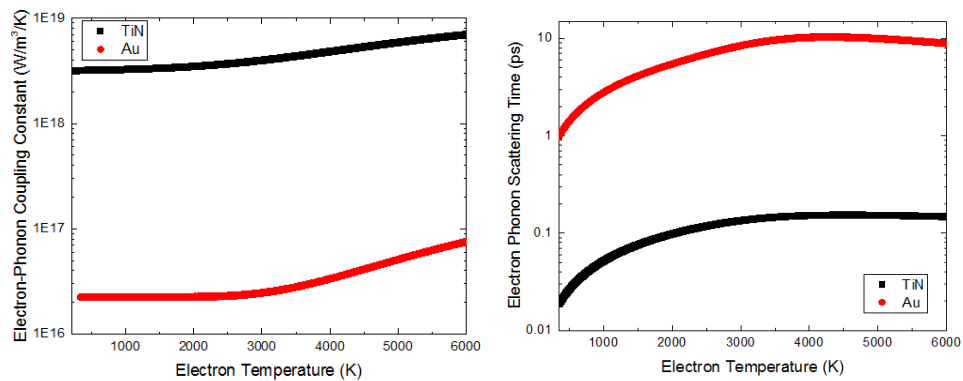

*Figure S3-2 | Calculated electron-phonon coupling and scattering time of TiN. a*, electron-phonon coupling constant of titanium nitride as a function of electron temperature displaying an increase over the entire range. *b,* electron-phonon scattering time as a function of electron temperature calculated from the parameters above.

[6] Brown, A. M., Sundararaman, R., Narang, P., Goddard, W. A. & Atwater, H. A. Ab initio phonon coupling and optical response of hot electrons in plasmonic metals. Phys. Rev. B **94**, 1–10 (2016).

S4 Pump-Power Invariance of Trapping Lifetimes

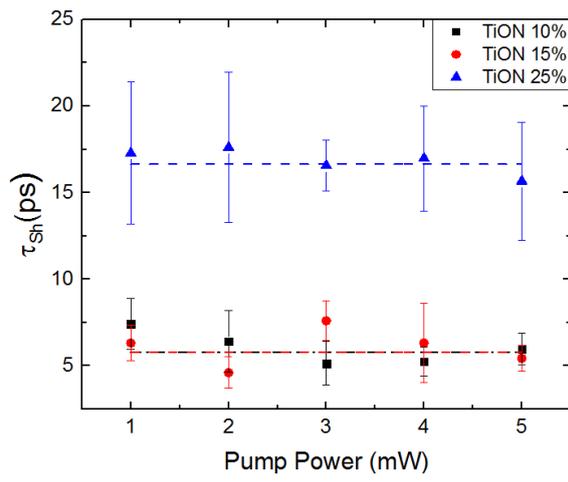
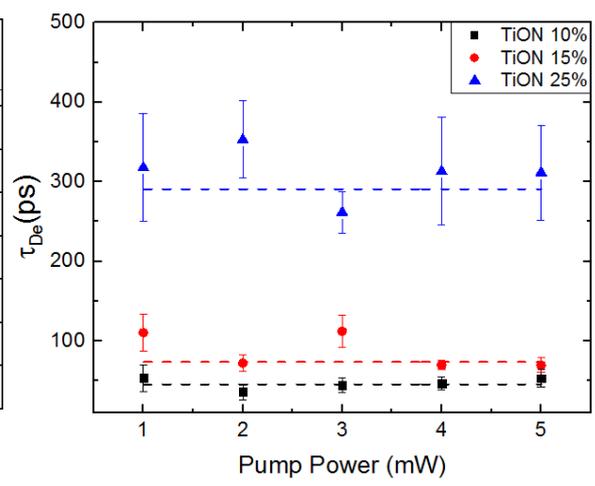